\documentclass[final,5p,times,twocolumn]{elsarticle}

\usepackage{graphicx}

\usepackage{amsmath}%
\usepackage{amsfonts}%
\usepackage{amssymb}%

\journal{Solid State Communications}

\begin{document}

\begin{frontmatter}



\title{Reply to ``On the cutoff parameter in the
translation-invariant theory of the strong coupling polaron''}


\author[1,2]{S. N. Klimin}
\author[1,3]{J. T. Devreese\corref{cor1}}
\cortext[cor1]{Phone: +32-3-2652485 \quad Fax: +32-3-2653318}
\ead{jozef.devreese@ua.ac.be}
\address[1]{Theorie van Kwantumsystemen en Complexe Systemen (TQC), Universiteit
Antwerpen, Universiteitsplein 1, B-2610 Antwerpen, Belgium}
\address[2]{Department of Theoretical Physics, State University of
Moldova, MD-2009 Chisinau, Moldova}
\address[3]{COBRA, Eindhoven University of Technology, 5600 MB Eindhoven,
The Netherlands}

\begin{abstract}
The present work is a reply to the paper \cite{Lakhno2}. It is proven that the
argumentation of Ref. \cite{Lakhno2} is inconsistent. The variational
functional for the polaron ground state energy considered in Ref.
\cite{Lakhno2} contains an incomplete recoil energy. Since the variational
functional of Ref. \cite{Lakhno2} is incomplete, it is not proven to provide a
variational upper bound for the polaron ground-state energy. The same
conclusion follows also for the bipolaron ground-state energy.
\end{abstract}


\end{frontmatter}






Polarons and bipolarons are invoked in the study of polar materials,
including high-$T_{c}$ superconductors \cite{R2,Alexandrov,Alex1}. 
Rigorous variational methods (see, e. g., Ref. \cite{Verbist1991,Cataudella,Kleinert})
are important in this field, i.~a. because in the bipolaron mechanism
of superconductivity, the parameters of the superconducting state
and the critical temperature strongly depend on the bipolaron binding energy.

The work \cite{Lakhno2} is a reply to our comments \cite{Comments} on the
variational approach aimed at in Refs. \cite{L1,L2,Tulub}. In Ref. \cite{Comments}
we show that the strong-coupling expression for the bipolaron ground state
energy calculated in Refs. \cite{L1,L2} is not justified as a variational
upper bound.

It is suggested in Ref. \cite{Lakhno2} that a properly chosen cutoff for the
phonon momenta leads to correct variational polaron and bipolaron ground-state
energies in the strong-coupling limit. However, this conclusion is not valid,
because the recoil energy treated in Refs. \cite{L1,L2,Tulub} is
\emph{incomplete}, as we wrote in Ref. \cite{Comments}.

The complete polaron recoil energy within the approach of Ref. \cite{Tulub}
was found by Porsch and R\"{o}seler \cite{Roseler}. They showed that, when
imposing a cutoff for the phonon momentum, the polaron recoil energy $E_{R}$
consists of \emph{two parts}:%
\begin{equation}
E_{R}=E_{R}^{\left(  T\right)  }+\delta E_{R}^{\left(  PR\right)  },
\label{ER}%
\end{equation}
where $E_{R}^{\left(  T\right)  }$ is the recoil energy determined in Ref.
\cite{Tulub}, and the term $\delta E_{R}^{\left(  PR\right)  }$ is given by
Eq. (43) of Ref. \cite{Roseler}:%
\begin{equation}
\delta E_{R}^{\left(  PR\right)  }=\frac{3\hbar}{2}\left(  \Omega_{q_{0}%
}-\omega_{q_{0}}\right)  , \label{PR}%
\end{equation}
where $q_{0}$ is the cutoff value for the phonon momentum, $\omega_{q}%
=\omega_{0}+\frac{\hbar q^{2}}{2m}$ with $\omega_{0}$ the LO-phonon frequency,
and $\left\{  \Omega_{q}\right\}  $ are the frequency eigenvalues resulting
from the Bogoliubov-like canonical transformation for the phonon operators
(performed in Refs. \cite{Tulub,Roseler}).

It is stated in the reply \cite{Lakhno2} that the reasoning of Ref.
\cite{Comments} is \textquotedblleft based on the erroneous approach ... to
the strong coupling limit when the cutoff parameter is introduced in the
theory.\textquotedblright\ However, the argumentation of Ref. \cite{Lakhno2}
is related only to the term $E_{R}^{\left(  T\right)  }$, ignoring the Porsch
--- R\"{o}seler term $\delta E_{R}^{\left(  PR\right)  }$. In the present work
we treat the contribution to the recoil energy $\delta E_{R}^{\left(
PR\right)  }$ missed in Refs. \cite{Lakhno2,L1,L2,Tulub}.

The expression obtained in Ref. \cite{Roseler} for $\Omega_{q_{0}}$
reads\footnote{There is a misprint in Eq. (42) of Ref. \cite{Roseler}
corresponding to Eq. (\ref{W}) of the present work: the factor 3 in the
denominator is missing.}%
\begin{align}
\Omega_{q_{0}} &  =\left\{  \omega_{q_{0}}^{2}+\int_{0}^{1}d\eta\int
_{0}^{q_{0}}dq\frac{\hbar q^{4}f^{2}\left(  q\right)  \omega_{q}}{3\pi^{2}%
m}\right.  \nonumber\\
&  \left.  \times\frac{2\operatorname{Re}F\left(  \omega_{q}+i\delta\right)
+\left\vert F\left(  \omega_{q}+i\delta\right)  \right\vert ^{2}}{\left\vert
1+F\left(  \omega_{q}+i\delta\right)  \right\vert ^{2}}\right\}
^{1/2}\label{W}%
\end{align}
with the function%
\begin{equation}
F\left(  z\right)  =\eta\frac{\hbar}{6\pi^{2}m}\int_{0}^{q_{0}}dq~q^{4}%
f^{2}\left(  q\right)  \left(  \frac{1}{\omega_{q}+z}+\frac{1}{\omega_{q}%
-z}\right)  .\label{F}%
\end{equation}
Here, $f\left(  q\right)  $ are variational functions. In Ref. \cite{Tulub},
they are chosen as%
\begin{equation}
f\left(  q\right)  =-\frac{V_{q}}{\hbar\omega_{0}}\exp\left(  -\frac{q^{2}%
}{2a^{2}}\right)  ,\label{fq}%
\end{equation}
with the variational parameter $a$ and the amplitudes of the electron-phonon
interaction $V_{q}$.

In Fig. 1, we plot the complete recoil energy $E_{R}$ and the contributions
$E_{R}^{\left(  T\right)  },$ $\delta E_{R}^{\left(  PR\right)  }$ as a
function of $\alpha$ for $q_{0}=8$ and $a=4$ (measured in units of
$\sqrt{\frac{m\omega_{0}}{\hbar}}$). The arrow indicates the value of the
coupling constant%
\begin{equation}
\alpha_{c}=\sqrt{2\pi}\frac{q_{0}^{4}}{a^{5}},\label{qc}%
\end{equation}
at which the steep maximum of the integrand in $E_{R}^{\left(  T\right)  }$
(mentioned in Ref. \cite{Tulub}) crosses the cutoff boundary.

For sufficiently small $\alpha$, the Tulub's recoil energy $E_{R}^{\left(
T\right)  }$ dominates, and $\delta E_{R}^{\left(  PR\right)  }$ is negligibly
small. When $\alpha$ increases (keeping other parameters constant),
$E_{R}^{\left(  T\right)  }$ tends to a finite value, while $\delta
E_{R}^{\left(  PR\right)  }$ monotonically increases.%

\begin{figure}
[h]
\begin{center}
\includegraphics[
height=2.3698in,
width=3.0801in
]%
{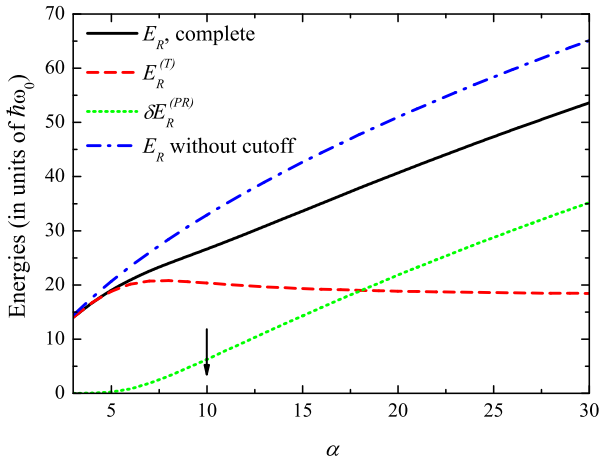}%
\caption{The recoil energy $E_{R}$ (solid black curve), the contributions
$E_{R}^{\left(  T\right)  }$ (dashed red curve) and $\delta E_{R}^{\left(
PR\right)  }$ (dotted green curve) as a function of $\alpha$ for $q_{0}=8$ and
$a=4$. The dot-dashed blue curve is the recoil energy without cutoff
\cite{Comments}. The arrow indicates the value $\alpha_{c}$ at which the peak
discussed in Ref. \cite{Tulub} passes the cutoff boundary.}%
\end{center}
\end{figure}

The analytic asymptotics for the Tulub and Porsch --- R\"{o}seler terms of the
polaron recoil energy in the strong-coupling limit gives us the results%
\begin{equation}
\left.  E_{R}^{\left(  T\right)  }\right\vert _{\alpha\gg1}\rightarrow\frac
{3}{16}a^{2}\left(  1+Q_{\infty}\right)  ,\quad Q_{\infty}=5.75\ldots
\label{SC0}%
\end{equation}
(as in Ref. \cite{Tulub}), and%
\begin{equation}
\left.  \delta E_{R}^{\left(  PR\right)  }\right\vert _{\alpha\gg1}%
\rightarrow\frac{3}{2}\left(  \frac{\sqrt{2}}{8\sqrt{\pi}}\alpha a^{5}\right)
^{1/2},\label{SC1}%
\end{equation}
which remarkably coincides with the asymptotic strong-coupling expression from
Ref. \cite{Comments} for the recoil energy without a cutoff. For an increasing
$\alpha$, the optimal value of the variational parameter $a$ increases. Thus
the Porsch --- R\"{o}seler contribution dominates in the recoil energy in the
strong-coupling regime, while the Tulub contribution $E_{R}^{\left(  T\right)
}$ constitutes only a residual part of the recoil energy. Moreover, the
analytic formula (\ref{SC1}) confirms the results of our previous treatment
\cite{Comments}. The same conclusion is valid for the bipolaron ground-state
energy, because the recoil contributions in the polaron and bipolaron problems
are structurally similar to each other.

In addition, there are logical inconsistencies in the argumentation of Ref.
\cite{Lakhno2} (which are of a secondary importance with respect to the
question discussed above). It is stated in Ref. \cite{Lakhno2} that
\textquotedblleft They ... used the asymptotics $q\left(  1/\lambda\right)
=2\sqrt{3\lambda}$ as the basis for their calculations of the polaron
energy.\textquotedblright\ However, the calculations in \cite{Comments} are
performed on the basis of the complete expression for the function $q\left(
1/\lambda\right)  $ given by Eqs. (2.11) to (2.12) of the work by Tulub
\cite{Tulub} rather than its strong-coupling asymptotics. Also it is written
in \cite{Lakhno2} that Tulub's choice of the variational functions $f\left(
q\right)  $ given by Eq. (\ref{fq}) is the best, while the choice of $f\left(
q\right)  $ in Ref. \cite{Comments} is the worst. However, the variational
functions used in Ref. \cite{Comments} are the same as those in Refs.
\cite{L1,L2,Tulub}.

\bigskip

In conclusion, when accounting for a phonon cutoff, the Porsch --- R\"{o}seler
contribution to the polaron and bipolaron recoil energy dominates in the
recoil energy in the strong-coupling regime calculated. This contribution is
missed in Refs. \cite{L1,L2,Tulub}. As a result, the variational functionals
for the polaron and bipolaron ground-state energies derived in Refs.
\cite{L1,L2,Tulub} are incomplete. Consequently, these variational functionals
are not rigorously proven upper bounds.

\bigskip
This work was supported by FWO-V projects G.0356.06, G.0370.09N, G.0180.09N,
G.0365.08, G.0115.12N, G.0119.12N, the WOG WO.033.09N (Belgium).


\begin{thebibliography}{9}
\bibitem {Lakhno2}V. D. Lakhno, \emph{arXiv:1206.3386v1}; Solid State
Communications (\emph{in press}).

\bibitem {R2} A. S. Alexandrov and J. T. Devreese,
\textit{Advances in Polaron Physics} (Berlin: Springer, 2009).

\bibitem {Alexandrov}A. S. Alexandrov, A. M. Bratkovsky, and N. F. Mott, Phys.
Rev. Lett. \textbf{72}, 1734 (1994).

\bibitem {Alex1}A. S. Alexandrov, Europhysics Letters, \textbf{95}, 27004 (2011).

\bibitem{Verbist1991} G. Verbist, F. M. Peeters, and J. T. Devreese, Phys. Rev. B \textbf{43}, 2712 (1991).

\bibitem{Cataudella} V. Cataudella, G. Iadonisi, and D. Ninno, Physica Scripta \textbf{T39}, 71 (1991).

\bibitem {Kleinert}H. Kleinert, in: \emph{Functional integration: basics and
applications}, pp. 93-95 (NATO Advanced Science Institutes Series B, Vol. 361, 1997).

\bibitem {Comments}S. N. Klimin and J. T. Devreese, Solid State Communications
\textbf{152}, 1601 (2012).

\bibitem {L1}V. D. Lakhno, JETP \textbf{110}, 811 (2010).

\bibitem {L2}V. D. Lakhno, Solid State Communications \textbf{152}, 621 (2012).

\bibitem {Tulub}A. V. Tulub, JETP \textbf{14}, 1828 (1961).

\bibitem {Roseler}M. Porsch and J. R\"{o}seler, Phys. Stat. Sol. (b)
\textbf{23}, 365 (1967).
\end{thebibliography}
\end{document}